\documentclass[fleqn,twoside]{article}
\usepackage{espcrc2}

\usepackage{graphicx}

\newcommand{\AmS}{{\protect\the\textfont2
  A\kern-.1667em\lower.5ex\hbox{M}\kern-.125emS}}
 \newcommand{\nc}{\newcommand}
 \nc{\mb}[1]{\makebox[#1]{}}
 \nc{\CC}{{\scriptscriptstyle CC}}
 \nc{\NC}{{\scriptscriptstyle NC}}
 \nc{\V}{{\rm v}}
 \nc{\W}{{\scriptscriptstyle W}}
 \nc{\X}{{\scriptscriptstyle X}}
 \nc{\Z}{{\scriptscriptstyle Z}}
 \nc{\CS}{{\scriptscriptstyle CS}}
 \nc{\DY}{{\scriptscriptstyle DY}}
 \nc{\PW}{{\scriptscriptstyle PW}}
 \nc{\SB}{{\scriptscriptstyle SB}}
 \nc{\CSV}{{\scriptscriptstyle CSV}}
 \nc{\GLS}{{\scriptscriptstyle GLS}}
 \nc{\CIB}{{\scriptscriptstyle CIB}}
 \nc{\PT}{{\scriptscriptstyle PT}}
 \nc{\IE}{{\it i.e.,\ }}
 \nc{\EG}{{\it e.g.,\ }}
 \nc{\EA}{{\it et al.\ }}
 \nc{\AH}{{\it ad hoc\ }}
 \nc{\CHPT}{{$\chi_{\PT}$\ }}
\nc{\st}{\scriptstyle}
\nc{\sst}{\scriptscriptstyle}
\nc{\mco}{\multicolumn}
\nc{\epp}{\epsilon^{\prime}}
\nc{\vep}{\varepsilon}
\nc{\ra}{\rightarrow}
\nc{\ppg}{\pi^+\pi^-\gamma}
\nc{\nuN}{{\nu N_0}}
\nc{\nubN}{{\overline{\nu} N_0}}
\nc{\snuNC}{{\langle \sigma^{\nuN}_{\NC}\rangle }}
\nc{\snubNC}{{\langle \sigma^{\nubN}_{\NC}\rangle }}
\nc{\snuCC}{{\langle \sigma^{\nuN}_{\CC}\rangle }}
\nc{\snubCC}{{\langle \sigma^{\nubN}_{\CC}\rangle }}
\nc{\Rnu}{{R^{\nu}}}
\nc{\Rnub}{{R^{\overline{\nu}}}}
\nc{\sintW}{{\sin^2 \theta_{\W} }}
\nc{\vp}{{\bf p}}
\nc{\rz}{{\rho_0^2}}
\nc{\ko}{K^0}
\nc{\kb}{\bar{K^0}}
\nc{\al}{\alpha}
\nc{\ab}{\bar{\alpha}}
\newcommand{\be}{\begin{equation}}
\newcommand{\ee}{\end{equation}}
\newcommand{\bea}{\begin{eqnarray}}
\newcommand{\eea}{\end{eqnarray}}
\title{Possible Explanations for The NuTeV Weinberg Angle  
Measurement}

\author{J.T. Londergan\address{Dept.\ of Physics, Indiana University \\ 
        Bloomington, IN, 47408, USA.}%
        \thanks{Research sponsored in part by US NSF under grant 
	NSF PHY0244822}
	}
       
\begin{document}

\begin{abstract}
The NuTeV collaboration has made an independent determination of the Weinberg 
angle by measuring charged- and neutral-current cross sections 
from neutrino and antineutrino DIS on iron. Their value 
differs by 3 standard deviations from that obtained from measurements 
at the $Z$ pole. We review this experiment and assess various 
possible explanations for this result, both within the Standard Model 
(``old physics''), and outside the Standard Model (``new physics'').   
\vspace{1pc}
\end{abstract}

\maketitle

\section{The NuTeV Experiment}

The NuTeV collaboration \cite{NuTeV} has measured total cross sections 
for neutrino and antineutrino scattering. 800 GeV protons at FNAL 
impinged on a BeO target, producing charged mesons; dipoles selected 
mesons of a particular charge, which subsequently decayed to muons and  
produced either neutrinos or antineutrinos. These were focused onto 
the NuTeV detector, an 18-m long, 690-ton long set of steel plates 
interspersed with liquid scintillator and drift chambers. Charged 
current (CC) and neutral-current (NC) interactions were selected according 
to the length 
of the subsequent track and the amount of missing energy. Events  
were selected using the criterion that the visible energy in the 
calorimeter satisfied the relation $20 < E_{vis} < 180$ GeV, and 
that the event vertex was contained within the fiducial volume. This 
resulted in a data sample of 1.62 million $\nu$'s, and 0.351 million 
$\bar{\nu}$'s. These cross sections were used to extract an independent 
value for the Weinberg angle, using a procedure suggested initially 
by Paschos and Wolfenstein \cite{Pas73}, who derived the relationship 
\bea
 R^{\PW} &\equiv& { \snuNC - \snubNC  \over 
 \rz\left( \snuCC - \snubCC \right)} = {1\over 2} - \sintW \nonumber \\ 
 &=& {\Rnu - \Rnub  \over 1 - r\, \Rnu} ~.  
\label{eq:PasW} 
\eea 
In Eq.\ {\ref{eq:PasW}, $\snuNC$ is the total NC inclusive 
cross section for neutrinos on an isoscalar target.  The quantity 
$\rho_0 \equiv M_{\W}/(M_{\Z}\,\cos \theta_{\W})$ is one in the Standard 
Model. The Paschos-Wolfenstein (PW) ratio can also be written in terms of 
ratios of NC to CC cross sections for neutrinos 
and antineutrinos; in Eq.\ \ref{eq:PasW}, $\Rnu$ is the NC/CC ratio  
of total cross sections induced by 
neutrinos on an isoscalar target, and $r$ is the ratio of CC  
cross sections for antineutrinos to CC cross sections 
induced by neutrinos.  

The NuTeV group obtained the neutral/charged ratios $\Rnu = 0.3916 \pm 
0.0007$ and $\Rnub = 0.4050 \pm 0.0016$, from which they extracted 
$\sintW = 0.2277 \pm 0.0013 ~(stat) \pm 0.0009 ~(syst)$.  This 
value is three standard deviations above the measured value for 
the Weinberg angle obtained from electroweak (EW) processes near the $Z$ pole, 
$\sintW = 0.2227 \pm 0.00037$ \cite{EM00}. This result was both striking 
and unexpected. It corresponds to a $1.2$\% decrease in the 
left-handed coupling $g_{\scriptscriptstyle{L}}^2$ of hadrons to the EW 
current. Derivation 
of the Paschos-Wolfenstein result contains a large number of assumptions: 
Among other things, it requires an isoscalar target; neglects contributions 
from quark masses; assumes isospin symmetry in the parton distribution 
functions (PDF's); neglects nuclear effects in PDFs; and neglects 
any contributions outside the Standard Model. Here we will review 
potential contributions from a number of these sources, and will 
suggest the most promising effects that might reconcile the NuTeV 
measurement with the extremely precise measurements near the $Z$ pole. 

\section{Contributions from Physics Outside the Standard Model} 

Davidson \EA have made an extensive review of possible contributions 
to the NuTeV measurement 
from ``New Physics,'' \IE physics beyond the Standard Model 
\cite{Dav03}. It is quite difficult to find new particles that contribute 
to the NuTeV neutrino measurement, but do not change other measured 
quantities from their experimental limits. This is due to the fact that 
several observables measured in EM experiments at the $Z$ pole 
are known to exceptional precision, in several cases at or near the 0.1\% 
level. It is hard to construct scenarios that will change the Weinberg 
angle measured in neutrino scattering by 1\%, without altering other 
observables by comparable amounts. To change the NuTeV neutrino result 
by the required amount without violating the constraints at the $Z$ pole, 
one has to resort to so-called 
``designer particles,'' new particles whose number, masses and couplings 
are delicately adjusted to fit all existing data. 

\begin{figure}[t]
\includegraphics[width=2.55in]{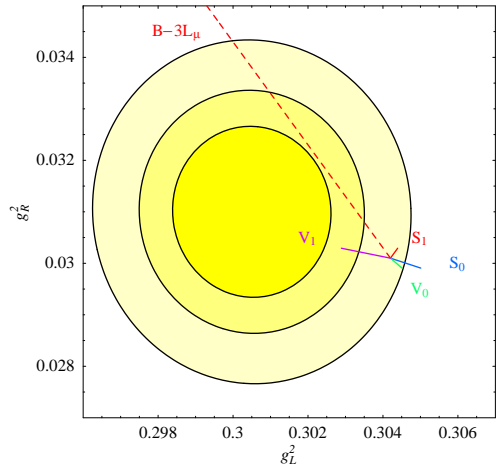}
\caption{Deviations from Standard Model for various types of new 
particles, without conflicting by more than $1\sigma$ with existing 
EW bounds, from Ref.\ \protect{\cite{Dav03}}, plotted vs.\ couplings of 
hadrons to EW currents $g_L^2$ and $g_R^2$. Circles represent standard 
deviations from NuTeV result. Solid lines: various leptoquark 
couplings; dashed line: 
extra $B-3L_{\mu}$ gauge boson.  
\label{Fig:Gambino}}
\end{figure}

We review a few of these attempts. One of the more obvious candidates 
might be loop effects containing supersymmetric particles. In general, 
{\bf SUSY loops in MSSM} (minimal supersymmetric models)  
tend to have the wrong sign, and increase the size of the anomaly; 
other SUSY candidates violate existing constraints. {\bf Oblique 
corrections}, arising from high-mass particles that couple only to vector 
bosons, have parameters that are 
tightly constrained by EW data, and they are probably unable to 
resolve the NuTeV anomaly.  A high-mass {\bf extra Z' vector boson} 
(or bosons), unmixed with the known $Z$ boson, might be a possibility. 
Davidson \EA found that the most likely possibility was a $Z'$ satisfying 
the gauge symmetry $B - 3L_{\mu}$, where $B$ is the baryon number and 
$L_{\mu}$ the lepton number. A gauge boson with precisely defined mass 
and couplings has the potential to remove the NuTeV anomaly;  
however, the couplings must be chosen to avoid running into trouble with 
the very precise muon $g-2$ measurements.   
{\bf Leptoquarks} are scalar or vector bosons that couple 
to both leptons and quarks. Assuming that the couplings conserve 
baryon and lepton number, it was found that only triplet 
$SU(2)_{\scriptscriptstyle{L}}$ 
leptoquarks, with carefully chosen mass splittings between the spin 
triplet members, could explain the NuTeV effect. If this explanation 
is true, it could be tested in experiments at the LHC. In Fig.\ 
\ref{Fig:Gambino}, we plot the contributions made by an extra $Z'$ with 
gauge symmetry $B - 3L_{\mu}$ (dashed curve), and from minimal 
leptoquarks (solid curves), vs.\ the left and right-handed couplings 
$g_L^2$ and $g_R^2$. The leptoquarks are differentiated according 
to the different $SU(2)_L$ symmetries; scalar singlet  
$S_0$ and and triplet $S_1$; and vector singlet $V_0$ and triplet 
$V_1$; only $V_1$ could remove the NuTeV anomaly.     

\section{Corrections Within the Standard Model} 

Because of the difficulty in satisfying the NuTeV result with physics 
outside the Standard Model, recent efforts to resolve this anomaly 
have focused on ``QCD Corrections,'' effects arising within the current 
Standard Model. We review the most promising effects here. In analyzing 
QCD corrections, we must remember an important feature of the NuTeV 
analysis. Because cuts and acceptance corrections are different for 
neutrinos and antineutrinos, the resulting ratios $\Rnu$ and $\Rnub$ 
cannot be combined simply, as in Eq.\ (\ref{eq:PasW}), to obtain the 
PW ratio and hence the Weinberg angle. The NuTeV group extracted the 
neutral to charged-current ratios, and the Weinberg angle, through a 
detailed Monte Carlo simulation of the reaction. This means that  
corrections to the Weinberg angle calculated from the PW ratio will not 
accurately predict the corresponding changes for the NuTeV experiment. 
For this purpose, the NuTeV group \cite{NuTeV2} has published functionals 
that provide the sensitivity of their observables to a given effect, 
\be
  \Delta {\cal E} = \int_0^1 \, F\left[ {\cal E}, \delta; x\right] 
  \, \delta(x) \, dx
\label{eq:Func}
\ee
Eq.\ \ref{eq:Func} gives the change in the extracted quantity 
${\cal E}$ resulting from the symmetry violating quantity $\delta(x)$.  
In the remainder of this paper, we will give equations showing the 
effect on the PW ratio resulting from various effects; however, the 
actual change in the Weinberg angle for the NuTeV measurement  
is obtained by integration using the relevant functional 
in Eq.\ (\ref{eq:Func}).  

\subsection{Nuclear Effects in Neutrino DIS} 

There is a correction to the Weinberg angle due to the fact that iron 
is not an isoscalar target. This correction is proportional to the 
number of excess neutrons in the target. It has been taken into account 
by the NuTeV group in extracting their value for the Weinberg angle. 
Although the correction is large, of the order -.008, it depends 
primarily on the momentum carried by valence quarks, a value  
known to within a couple percent; hence this correction should be 
well under control. Another significant contribution arises from 
radiative corrections; these occur only for the charged-current events, 
and involve coupling of soft photons to the final muon line. The 
NuTeV group used radiative corrections calculated by Bardin and 
Dokuchaeva \cite{Bar86}. Recently, Diener, Dittmaier and Hollik 
\cite{Die04} have re-calculated the radiative corrections. Their 
program has not yet been included in a detailed analysis of the 
$\nu$ cross sections; however, they 
suggest that the NuTeV estimate of the uncertainty due to radiative 
corrections ($\delta \sintW = 0.00011$) may be overly optimistic.  

There should be nuclear corrections, since  NuTeV measurements were 
carried out on an iron target. For charged 
leptons, one has measured shadowing corrections 
at small $x$, EMC-type corrections at intermediate $x$, and Fermi 
motion corrections at large $x$.  
Miller and Thomas \cite{Mil02} suggested that shadowing effects  
might make a significant correction to the NuTeV result. Their argument 
is that shadowing for neutrinos (virtual $W$ or $Z$) could be significantly 
different from shadowing for muons (virtual photons) 
\cite{Bor98}. The NuTeV group counters that shadowing occurs at values of 
$Q^2$ significantly lower than the $Q^2$ in their experiment. They 
also claim that the shadowing corrections should cancel when one 
subtracts neutrino cross sections from antineutrino reactions; 
furthermore, shadowing should affect the ratio $\Rnub$ more than 
$\Rnu$, while in their experiment $\Rnub$ is  
identical to the Standard Model prediction, while $\Rnu$ is $3\sigma$  
below the Standard Model value. 

In a recent paper, Brodsky 
\EA \cite{Bro04} calculate shadowing and antishadowing corrections 
for neutrino-nucleus DIS. They include Pomeron and Odderon contributions 
from multigluon exchange, in addition to Reggeon multiquark exchange. 
In their model CC and NC reactions are affected differently, as are 
$\nu$ and $\bar{\nu}$ processes. They estimate that shadowing effects 
can account for roughly 20\% of the NuTeV anomaly; their calculations 
predict significantly larger corrections for $\Rnub$ than for $\Rnu$. 
Kumano \cite{Kum02} estimated nuclear modifications 
to the PW ratio, and Hirai \EA \cite{Hir04} are 
currently extracting nuclear PDFs by fitting a wide range of 
data for charged lepton DIS and Drell-Yan reactions. The NuTeV 
group has applied nuclear corrections, under the assumption  
that nuclear modifications are identical for charged-lepton and 
$\nu$ DIS (one expects differences in the shadowing 
region, and it is not firmly established that EMC-type 
modifications are the same for charged leptons and neutrinos).    

\subsection{Contributions from Isospin Violation} 

Charge symmetry is an approximate symmetry in parton distributions, 
involving rotation of $180^\circ$ about the 
$``2''$ axis in isospin space; it corresponds to interchange 
of protons and neutrons (simultaneously interchanging up and down 
quarks). In nuclear physics, charge symmetry 
is respected to a high degree of precision, generally at less than 
the 1\% level \cite{Miller,Henley}. At present, there is no direct 
evidence for charge symmetry violation (CSV) in high energy experiments 
\cite{Lon98}, and until recently all phenomenological analyses assumed 
charge symmetry for PDFs. We define CSV PDFs as $\delta d_{\V}(x) \equiv 
d_{\V}^p(x) - u_{\V}^n(x)$ and $\delta u_{\V}(x) \equiv u_{\V}^p(x) - 
d_{\V}^n(x)$. The correction to the PW ratio from parton 
charge symmetry violation has the form 
\bea 
\Delta R^{\PW}_{\CSV} &\approx& \left[ 1- {7\over 3}\sintW \right] 
 {\delta U_{\V} - \delta D_{\V} \over 2(U_{\V} + D_{\V})} 
  \nonumber \\ Q_{\V} &\equiv& \int_0^1 x\,[q(x)- \bar{q}(x)]\,dx 
\label{eq:PWcorr}
\eea
In Eq.\ (\ref{eq:PWcorr}), we have for simplicity left out a small 
QCD radiative correction proportional to $\alpha_s$ \cite{Dav03}. 
Only valence PDFs contribute to the PW ratio. 

Recently, the MRST group \cite{MRST03} have included isospin violation 
in a phenomenological evaluation of PDFs. They carried out a global 
fit of PDFs using a wide variety of high energy data. They   
chose a specific model for valence quark charge symmetry violating PDFs. 
The integral over $x$ of the valence CSV PDFs $\delta d_{\V}(x)$ and 
$\delta u_{\V}(x)$ must be zero, otherwise it would not conserve the 
total number of valence quarks in the neutron. MRST constructed a function 
that automatically satisfied this quark normalization condition, 
\bea 
 \delta u_{\V}(x) &=& - \delta d_{\V}(x) = \kappa f(x) \nonumber \\ 
  f(x) &=& (1-x)^4 x^{-0.5}\, (x - .0909)  \, .
\label{eq:CSVmrst}
\eea
At both small and large $x, f(x)$ 
has the same form as the MRST valence quark distributions, and the 
first moment of $f(x)$ is zero. The coefficient $\kappa$ was varied in a 
global fit to high energy data. MRST obtained a shallow $\chi^2$ minimum 
about the value $\kappa = -0.2$, and the 90\% confidence level 
for the range $-0.8 \le \kappa \le +0.65$.  

The MRST group also included sea quark CSV in a global fit. 
Valence quark CSV makes a substantially larger 
contribution than sea quark CSV to the extraction of the Weinberg angle from 
neutrino DIS.  Using the sea-quark CSV and the best-fit value for valence 
quark CSV obtained by the MRST group, would remove roughly 
$1/3$ of the NuTeV anomaly. Note that the form chosen by MRST, Eq.\ 
(\ref{eq:CSVmrst}), insures that $\delta u_{\V}$ and $\delta d_{\V}$ 
will have opposite sign, and hence will add coherently in the correction 
to the PW ratio, Eq.\ (\ref{eq:PWcorr}). The value $\kappa = -0.6$, within the 
90\% confidence limit found by MRST, would completely remove the 
NuTeV anomaly, while the value $\kappa = +0.6$ would double the 
discrepancy.  The MRST results show that isospin 
violating PDFs are able to completely remove the NuTeV anomaly in the 
Weinberg angle, or to make it twice as large, without serious disagreement  
with any of the data used to extract quark and gluon PDFs. 

\begin{figure}[t]
\vspace{-1.0cm}
\hspace{-1.2cm} 
\includegraphics[width=3.4in]{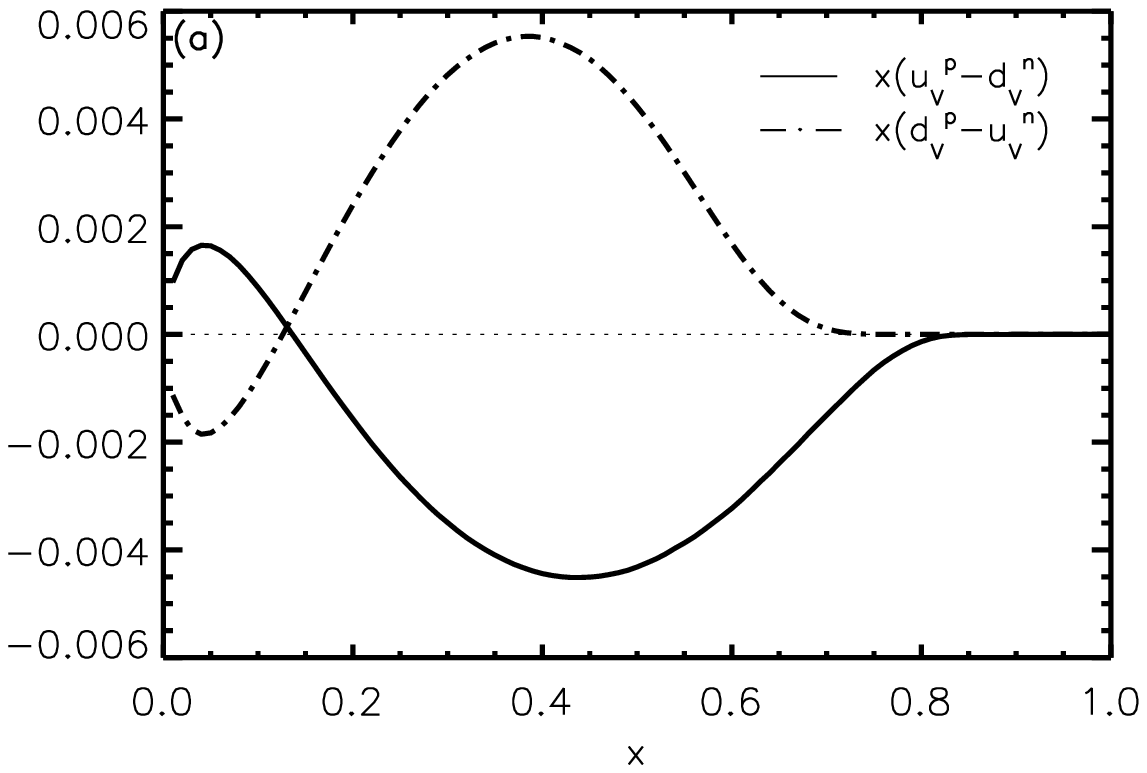}
\includegraphics[width=2.2in]{MRSTcsv.eps}
\caption{Above: theoretical 
CSV PDFs by Rodionov \EA, Ref. \protect\cite{Rod94}. Solid line: 
$x\delta u_{\V}$; dash-dot line: $x\delta d_{\V}$. Below: MRST valence 
CSV PDF from Ref. \protect\cite{MRST03}, corresponding 
to best fit value 
$\kappa = -0.2$ in Eq.\ (\protect\ref{eq:CSVmrst}). Solid curve: 
$x\delta d_{\V}(x)$; dashed curve: $x\delta u_{\V}(x)$. 
\label{Fig:MRSTfx}}
\end{figure}

We can compare the phenomenological fits of MRST to 
theoretical estimates of 
valence parton CSV by Sather \cite{Sat92} and Rodionov 
\EA \cite{Rod94}. By estimating the dependence of valence PDFs on 
the quark and nucleon masses, Sather obtained an analytic approximation 
to valence quark PDFs; the second moment of these distributions gave 
the results 
\be
\delta U_{\V} = {\delta M\over M}[U_{\V} -2]; \hspace{0.6cm} 
 \delta D_{\V} = {\delta M\over M}D_{\V} + {\delta m\over M}
\label{eq:Satmoment}
\ee
In Eq.\ (\ref{eq:Satmoment}), $\delta m = m_d - m_u \sim 4$ MeV is the 
quark mass difference, while $\delta M = M_n - M_p = 1.3$ MeV is the 
n-p mass difference. Eq.\ (\ref{eq:Satmoment}) suggests that the 
second moments $\delta U_{\V}$ and $\delta D_{\V}$ depend only upon the 
total momentum carried by valence quarks in the nucleon, and the 
n-p and up-down quark mass differences. It also suggests that 
$\delta U_{\V}$ should be negative and $\delta D_{\V}$ positive; since 
the PW ratio is proportional to the difference of these quantities, 
they will add coherently in the correction to the PW ratio.    
Rodionov \EA calculated the valence CSV distribution 
expressions for valence quark distributions obtained from quark models. 
In both cases, the resulting CSV corrections tend to decrease the 
size of the NuTeV anomaly, and they remove roughly 30\%, or $1\sigma$   
of the discrepancy in the Weinberg angle.    
In Fig.\ \ref{Fig:MRSTfx} we compare the valence quark CSV PDFs for the 
MRST best fit value $\kappa = -0.2$ with the theoretical CSV distributions 
of Rodionov {\it et al.}. The two distributions agree very well. 
However, within the 90\% confidence level the phenomenological CSV 
distributions could be four times as large as the theoretical PDFs, or three 
times as big with the opposite sign. 

\subsection{Strange Quark Contribution} 

The contribution from strange 
quarks has the form 
\be 
\Delta R^{\PW}_{s} \approx \left[ 1- {7\over 3}\sintW \right] 
 {-S_{\V}  \over 2(U_{\V} + D_{\V})} \,.  
\label{eq:stcorr}
\ee
It depends on the quantity $S_{\V} \equiv \int_0^1 x[s(x)- 
\bar{s}(x)]dx$, the asymmetry in total momentum carried by strange 
quarks and antiquarks. The most precise measurements 
come from opposite sign dimuons produced from reactions induced by 
neutrinos or antineutrinos; the former (latter) is sensitive to the 
strange (antistrange) quark distributions. This has been measured by 
both CCFR \cite{Baz93} and NuTeV \cite{Gon01} groups. The CTEQ group 
\cite{Kret04} has analyzed this in a global fit that allowed 
the possibility of $s \ne \bar{s}$. They find a positive value for 
$S_{\V}$, which would decrease the Weinberg anomaly; the most likely 
value is roughly a 30\% decrease in the NuTeV anomaly for the Weinberg 
angle; in their analysis, the strange contribution alone is unable to 
remove the NuTeV anomaly. The NuTeV group had previously reported a best 
fit to their dimuon cross sections with a slightly negative value 
$S_{\V} = -.0027 \pm 0.0013$ 
\cite{NuTeV2}; this would increase the magnitude of the NuTeV anomaly. 
In Fig.\ \ref{Fig:squark} we plot the results of the CTEQ and NuTeV 
analyses of strange quark distributions; we show both $s-\bar{s}$, and 
the second moment $x(s-\bar{s})$ (multiplied by 5). 

\begin{figure}
\includegraphics[width=2.5in]{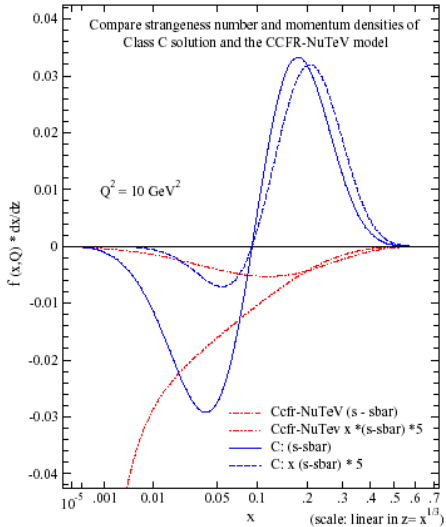}
\vspace{-0.5cm}
\caption{Positive values at large $x$: $s-\bar{s}$ and $5x(s-\bar{s})$, 
from CTEQ global fit, Ref.\ \protect{\cite{Kret04}}. Negative values: the 
same quantities calculated in earlier CCFR/NuTeV analysis of strange quark 
distributions, Ref.\ \protect{\cite{Gon01}}.
\label{Fig:squark}}
\end{figure}

Since the initial NuTeV and CTEQ results, the groups have worked together 
to resolve the differences in their analyses. Initially, there were 
rather substantial differences. At present, a few 
differences still persist: CTEQ uses a somewhat lower charm mass  
(necessitated by HERA data in their global fit); and the two groups 
use slightly different acceptance corrections and fragmentation functions. 
In their 
global fit, the CTEQ group calculates dimuon production using an LO 
fit, while NuTeV uses NLO calculations throughout. The CTEQ 
group enforced the condition that the proton have 
zero net strangeness $\int_0^1 [s(x)-\bar{s}(x)]dx =0$, while at 
least initially NuTeV did not (this is apparent in Fig.\ 
\ref{Fig:squark}). The two groups continue to obtain opposite 
overall sign for the strange quark contribution to the Weinberg angle. 
It is not clear why this is the case; although the CTEQ group carries out 
a global fit while NuTeV specifically fits the CCFR/NuTeV dimuon production 
cross sections, determination of the strange quark PDFs in the global fit 
is dominated by the dimuon data. Catani \EA \cite{Cat04} have recently 
shown that, at NNLO order, it is possible to perturbatively generate nonzero 
values for $S_{\V}$, even if it is zero at the starting scale (this 
is not possible in either LO or NLO). They generate a small negative 
value for $S_{\V}$ from perturbative three loop contributions. 

\section{Conclusions\label{Sec:Concl}} 

It has proved quite difficult to find physics outside the Standard Model 
that can reconcile the NuTeV measurement of the Weinberg angle, while 
agreeing with the extremely precise EW data measured at the 
$Z$ pole. Recent attention has focused on QCD corrections. Of the effects 
we reviewed, the most viable candidate 
would be corrections from isospin violating PDFs. The MRST group has shown 
that CSV effects are capable of removing 100\% of the NuTeV anomaly, 
without disagreeing seriously with any high energy data. If correct, these 
parton CSV effects would be roughly three times theoretical predictions; 
effects of this magnitude should be visible at the few percent level in 
future experiments. Another possible explanation for the NuTeV anomaly 
would be a strange quark momentum asymmetry. There is currently disagreement 
about both the sign and magnitude of this contribution; the CTEQ group 
estimates that $s$ quark effects should remove roughly $1/3$ of the NuTeV 
anomaly, while the NuTeV analysis of the same data suggests a small 
contribution from strange quarks that is likely to increase the anomaly. 
Another potential contribution arises from nuclear effects in neutrino 
scattering. Recent calculations suggest that nuclear effects could 
remove part of the NuTeV anomaly. One possible scenario is that the 
discrepancy in the Weinberg angle could be explained by small contributions 
from several QCD effects. However, it should be stressed that 
for all of these processes, the sign of the contribution to the NuTeV 
experiment has not been established beyond doubt, so we cannot say 
with certainty that these effects reduce the discrepancy 
in the Weinberg angle.

Theoretical work cited here was carried out  
with A.W. Thomas and other collaborators at CSSM, Adelaide. Thanks to  
G.P. Zeller and K. McFarland for useful comments regarding the NuTeV 
measurements, S. Kretzer and F. Olness for discussions regarding 
the CTEQ analysis, and R. Thorne for information regarding the 
MRST global PDFs.    



\begin{thebibliography}{99}  

\bibitem{NuTeV} NuTeV Collaboration, G.P. Zeller \EA, Phys. Rev. Lett.  
        {\bf 88}, 091802 (2002).   

\bibitem{Pas73} E. A. Paschos and L. Wolfenstein, Phys. Rev. {\bf D7}, 
                          91, (1973). 

\bibitem{EM00} D. Abbaneo \EA, hep-ex/0112021.  

\bibitem{Dav03}  S. Davidson, S. Forte, P. Gambino, N. Rius, and 
	A. Strumia, JHEP {\bf 202}, 037 (2002).

\bibitem{NuTeV2} NuTeV Collaboration, G.P. Zeller \EA, Phys. Rev.   
        {\bf D65}, 111103 (2002).   

\bibitem{Bar86} D. Yu. Bardin and V.A. Dokuchaeva, report 
	JINR-E2-86-260.  

\bibitem{Die04} K-P.O. Diener, S. Dittmaier and W. Hollik,  
              Phys. Rev. D{\bf 69}, 073005 (2004).
   
\bibitem{Mil02}  G.A. Miller and A.W. Thomas, arXiv:hep-ex/0204007.

\bibitem{Bor98}  C. Boros, J. T. Londergan and A. W. Thomas,
              Phys. Rev. D{\bf 58}, 114030 (1998).

\bibitem{Bro04}  S.J. Brodsky, I. Schmidt and J-J. Yang, 
	arXiv:hep-ph/0409279.

\bibitem{Kum02}  S. Kumano, Phys.\ Rev.\ D{\bf 66}, 111301 (2002).

\bibitem{Hir04} M. Hirai, S. Kumano and T-H. Nagai,   
              arXiv:hep-ph/0408023; hep-ph/0408135.
   
\bibitem{Miller} G. A. Miller, B. M. K. Nefkens and I. Slaus, Phys. Rep. 
       {\bf 194},1  (1990). 

\bibitem{Henley} E. M. Henley and G. A. Miller in {\it Mesons 
         in Nuclei}, eds M. Rho and D. H. Wilkinson 
         (North-Holland, Amsterdam 1979). 

\bibitem{Lon98} J. T. Londergan and A. W. Thomas, 
          in {\it Progress in Particle and Nuclear Physics}, 
           Volume 41, p.\ 49,   
          ed.\ A. Faessler (Elsevier Science, Amsterdam, 1998).  

\bibitem{MRST03}  MRST Collaboration, A.D. Martin \EA, 
	arXiv:hep-ph/0308087.

\bibitem{Sat92} E. Sather, Phys. Lett. {\bf B274}, 433 (1992).

\bibitem{Rod94} E. Rodionov, A. W. Thomas and J. T. Londergan,
              Mod. Phys. Lett. {\bf A9}, 1799 (1994).

\bibitem{Baz93} CCFR Collaboration, A.O. Bazarko \EA, Z. Phys.\ 
	C{\bf 65}, 189 (1995). 

\bibitem{Gon01} NuTeV Collaboration, M. Goncharov \EA, Phys.\ Rev.\ 
	D{\bf 64}, 112006 (2001). 

\bibitem{Kret04} S. Kretzer, arXiv:hep-ph/0405221; F. Olness \EA 
	arXiv:hep-ph/0312323; S. Kretzer \EA arXiv:hep-ph/0312322.   

\bibitem{Cat04} S. Catani \EA, arXiv:hep-ph/0404240; G. Rodrigo \EA, 
	arXiv:hep-ph/0406338.   

\end{thebibliography}
\end{document}